\documentclass[aps,prl,preprint,superscriptaddress]{revtex4-2}

\usepackage[T1]{fontenc}

\usepackage{amsfonts}
\usepackage{amsmath}
\usepackage{upgreek}  
\usepackage{mathrsfs}
\usepackage{booktabs} 
\usepackage{csquotes}
\usepackage{blindtext}
\usepackage{epigraph}
\usepackage{wrapfig}
\usepackage{graphicx}
\usepackage{subcaption}
\usepackage{comment}

\renewcommand{\vec}[1]
{
{\mathbf #1}
}

\renewcommand{\Im}
{
{\mathfrak{Im}}
}
\newcommand{\ii}
{
{{\mathfrak{i}}} 
}

\newcommand{\fbr}[1]
{
{\left( #1 \right)} 
}

\newcommand{\xc}
{
_{\mathrm{xc}}
}
\newcommand{\KS}
{
_{\mathrm{KS}}
}

\newcommand{\Loss}{\mathcal{L}}

\newcommand{\vC}{v_{\textrm{C}}}

\begin{document}


\title{Dispersion and lifetimes of magnons in non-collinear magnets from time dependent density functional theory}


\author{David Eilmsteiner}
\affiliation{Institute for Theoretical Physics, Johannes Kepler University, Linz, Austria}
\affiliation{Hamburg University of Applied Sciences, Hamburg, Germany}
\author{Arthur Ernst}
\affiliation{Institute for Theoretical Physics, Johannes Kepler University, Linz, Austria}
\affiliation{Max Planck Institute of Microstructure Physics, Weinberg 2, 06120, Halle (Saale), Germany}
\affiliation{Donostia International Physics Center (DIPC), 20018 Donostia-San Sebasti\'{a}n, Spain}
\author{Pawe\l{} A. Buczek}
\email[]{pawel.buczek@haw-hamburg.de}
\affiliation{Hamburg University of Applied Sciences, Hamburg, Germany}


\date{\today}
\begin{abstract}
We investigate the spin dynamics of the non-collinear kagome triangular anti-ferromagnet Mn$_3$Rh using linear response time-dependent density functional theory.
To this end, we present a novel first principles computational scheme for the evaluation of the dynamical susceptibility based on the non-collinear KKR Green's functions method and a symbolic computer algebra.
This approach allows us 
to address the Landau decay of spin waves into non-collinear electron-hole Stoner pairs
being inaccessible to adiabatic methods. 
Our calculations reveal three distinct Goldstone modes dispersing linearly in the long-wavelength regime giving rise to the three magnon branches
and we discuss their non-trivial spatial polarizations.
The spin-waves turn out to be defined in the whole Brillouin zone but their Landau damping becomes substantial away from the zone's center.
Surprisingly, magnons of comparable momenta and energies can feature, depending on their chirality, considerably different attenuation, in some cases of predominantly resonant character.
We trace this effect to the interplay between the magnon eigenvectors and 
the intrinsically spin-polarized altermagnetic band structure and the resulting spectrum of non-collinear Stoner states.
\end{abstract}


\maketitle

\begin{figure}
	\includegraphics[width=0.9\linewidth,trim={60 140 120 140},clip]{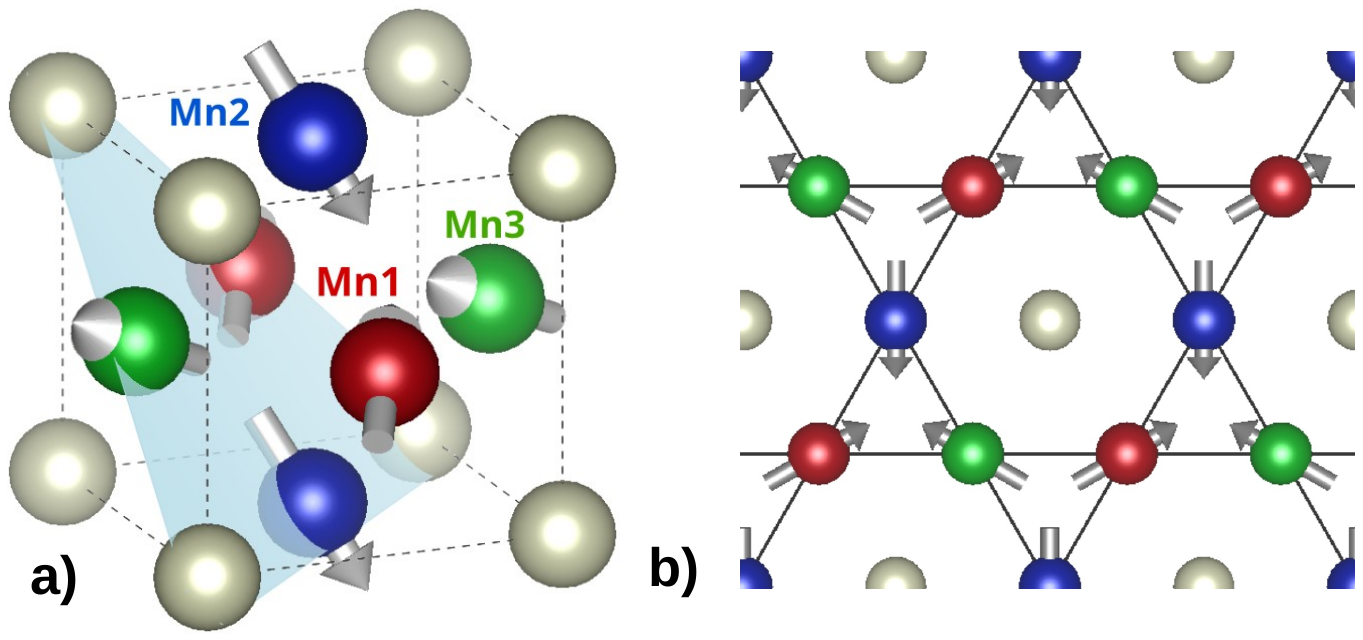}
	\caption{(a) Magnetic unit cell of the KAFM Mn$_3$Rh with the Mn atoms depicted in color depending on their orientation and the Rh atoms in grey. The (111) plane is marked in light blue. Arrows show the magnetic moment directions. (b) The top view on the (111) plane reveals the characteristic frustrated KAFM.}
	\label{fig:structure}
\end{figure}

\textit{Introduction - }
We witness spectacular progress in the area of magnetism inextricably linked to the advent of fascinating novel classes of functional materials. As prominent examples, skyrmionic matter \cite{nagaosaTopologicalPropertiesDynamics2013g}, altermagnets \cite{songAltermagnetsNewClass2025}, and non-collinear (NC) magnetic systems \cite{Rimmler2025} 
promise to be the stuff that new generations of spintronic computers will be made of. 
The topologically protected skyrmionic quasiparticles are envisioned for information processing and storage \cite{fertMagneticSkyrmionsAdvances2017,gubbiotti2025Roadmap3D2025}. Altermagnets blend the advantages of antiferromagnets (AFMs) and ferromagnets \cite{Mazin2022,cheongAltermagnetismNoncollinearSpins2024,mazinAltermagnetismThenNow2024,Maznichenko2024a,Sandratskii2025}, and 
their potential in magnonics and magneto-optics has been pointed out \cite{Smejkal-PRX12-040501}.
They feature an intrinsically spin-polarized band structure while exhibiting a vanishing net magnetic moment.
This remarkable feature is also found in another family of intricate magnets, the frustrated NC kagome AFMs (KAFMs) \cite{gurungNearlyPerfectSpin2024b,huSpinHallEdelstein2025}.
The magnetic non-collinearity abandons the single electron spin as a good quantum number, allowing the direction of magnetization to vary in space.
The existence of such systems is experimentally established, prominent examples being Mn$_3$X with X$\in[\text{Pt,Ir,Rh}]$ (cf. Fig.~\ref{fig:structure}) featuring the kagome magnetic ordering and an extraordinary spintronic application potential \cite{Rimmler2025}.
The beautiful symmetries of NC magnets (NCMs) \cite{Sandratskii1998,pradenasSpinFrameFieldTheory2024,Pradenas2025} and the puzzling appearance of this order close to quantum critical points \cite{zlotnikovAspectsTopologicalSuperconductivity2021,dahlbergSpinglassDynamicsExperiment2025} fuels the unquenched interest even more.
Apart from the ground state determination, a faithful description of magnetization dynamics, in particular the emergence of collective low energy spin-waves (magnons) lies at the the heart of the physical description of any magnetic system.
Spin excitations govern the thermodynamics of magnets and couple to electronic degrees of freedom \cite{Paischer2023}, modifying the band structure \cite{Mlynczak2019} and even allowing for the formation of unconventional Cooper-pairs \cite{Essenberger2016}.
Furthermore, any magnonic application requires a realistic description of magnons, specifically of their dispersion, the spatial form of modes (polarization), and, last but not least, their life-time.


The decay of magnons acquires a fascinating aspect in triangular AFMs.
The latter elude the linear spin-wave theory and their magnons undergo intrinsic damping due to two-boson interactions, even at absolute zero \cite{Chernyshev2009,Zhitomirsky2013}.
In addition, there are further fundamentally important magnon attenuation channels.
In imperfect crystals, disorder plays a pivotal role \cite{Paischer2021a}
and, in metallic magnets, the collective spin-waves decay into electron-hole pairs in the process of Landau damping.
In collinear magnets (CM), these pairs are termed Stoner excitations and involve electrons of opposite spins.
This simple picture must be non-trivially generalized for NCMs, which is one of the aims of this paper.
The Landau attenuation can become substantial but depends strongly on specific material and dimensionality \cite{Buczek2011a}, thus, requiring an \textit{ab initio} treatment.
On the other hand, it is completely neglected in the Heisenberg model-based theories \cite{leblancSpinWavesAnisotropic2014,leblancImpactFurtherrangeExchange2021}
but can be captured within the many-body perturbation theory \cite{Aryasetiawan1999,Mueller2016,Binci2025} or by resorting to the linear response time-dependent density functional theory (LRTDDFT)
\cite{%
Savrasov1990,%
Buczek2010d,%
Lounis2011,%
Skovhus2022}.
In both cases, the magnon energies and life-times are extracted from the singularities of the dynamic response function, the magnetic susceptibility.
Apart from the capturing of the Landau damping, the scheme avoids ambiguities associated with Heisenberg model mapping \cite{Binci2025} and constitutes the state-of-the-art theory of spin excitations.
Alas, its numerical and algorithmic complexity has precluded a widespread use, especially for complex and non-collinear systems.
In this paper, the LRTDDFT formalism is applied for the first time to NCMs
and we outline the non-trivial formal and computational extensions of our scheme necessary to treat the NC case.
Following it, we address the spin dynamics of extremely relevant family of kagome magnets with Mn$_3$Rh being its exemplary member.
Our calculations reveal three distinct Goldstone modes dispersing linearly in the long-wavelength regime.
We discuss their non-trivial polarization and proceed to an in-depth analysis of their dispersion and Landau damping governed by the intricacies of the non-collinear electronic band structure.

\begin{figure*}
	\includegraphics[width=0.95\textwidth,trim={160 0 60 0},clip]{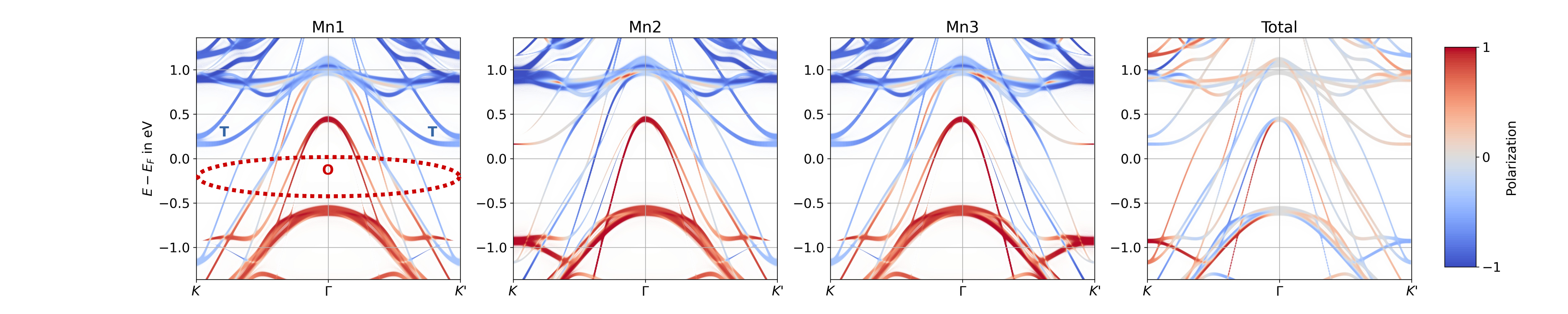}
	\caption{Contribution of the magnetic Mn atoms to the total non-collinear altermagnetic band structure of Mn$_3$Rh (left). The states' polarization is marked in color, with red (blue) signifying spin up (down) in the moment's local frame of reference. The intensity of the band signifies the Bloch amplitude of the state on the atom. Solid and dashed rings mark T and O states, respectively.  For the full total band structure (far right), the spin directions are relative to the frame of reference of Mn2.}
	\label{fig:bsf}
\end{figure*}

\textit{Method -}
We outline now the major features or our scheme \cite{Eilmsteiner2026}.
For periodic crystals, the true dynamical susceptibility $\chi^{ij}\fbr{\vec{q}, \vec{r}, \vec{r}', \omega}$ relates the charge and magnetization density response ($i,j=0,x,y,z$) to scalar and magnetic fields coupling to these densities \cite{Gross1985}. $\vec{q}$ resides in the first Brillouin zone
and $\vec{r}, \vec{r}'$ belong to the crystal's primitive cell allowing for the determination of the spatial shapes of the excitations.
The collective density modes (magnons, plasmons, etc.) are identified as eigenvectors of the anti-Hermitian part of the susceptibility, the loss matrix $\Loss(\omega) \equiv - \ii \fbr{\chi(\omega) - \chi^{\dagger}(\omega)}/2$.
In the virtue of the fluctuation-dissipation theorem, the corresponding eigenvalues yield the density of these excited states in the energy space.
$\chi$ is found upon solving the susceptibility Dyson equation \cite{Gross1985}
\begin{align}
  \chi = \chi\KS + \chi\KS (\vC + K\xc) \chi 
\label{eq:suscDysonEquation}
\end{align}
Above, the $\fbr{\vec{q},\omega}$ are suppressed for brevity and all quantities are assumed to be matrices representing the $\fbr{i\vec{r}, j\vec{r}'}$-dependence in a given basis which we choose to be atom centered for numerical expediency.
$\vC$ is the Coulomb interaction (\enquote{Hartree response}), and $K\xc$ is the so called exchange correlation kernel.
It describes the change of the effective Kohn-Sham field induced by the densities arising due to the external fields.

The exact form of $K\xc$ is unknown
and its determination is as involved as the solution of the many-body electronic problem itself.
In this work, we adopt the adiabatic local spin density approximation (ALSDA) \cite{Katsnelson2004}.
$\chi^{ij}\KS\fbr{\vec{q}, \vec{r}, \vec{r}', \omega}$ describes the response of the formally non-interacting Kohn-Sham system at ground state density and is determined by the electronic band structure.
It contains information about the particle-hole (Stoner) pairs involved in the Landau damping.
In the ALSDA, the $K\xc$ is approximated by its static component.
While this still allows us to describe the Landau damping through the frequency dependence of
$\chi\KS$, the  magnon-magnon induced decay in the LRTDDFT must formally enter through the non-adiabaticity of $K\xc$, an issue which will be addressed elsewhere.
$\chi\KS$ is found as a product of two Kohn-Sham Green's functions (KS GF)
\begin{equation}
  \chi^{ij}\KS \equiv
    \sigma^{i}_{\alpha\beta}
    G_{\beta\gamma}
    \sigma^{j}_{\gamma\delta}
    G_{\delta\alpha}
\label{eq:chiKS}
\end{equation}
where $\sigma^{i}$ stands for the Pauli matrix.
The GF is obtained using the Korringa-Kohn-Rostoker (KKR) method
\cite{%
lounisNoncollinearKorringaKohnRostokerGreen2005a,%
hoffmannMagneticElectronicProperties2020%
}.  
Although formally similar to the scheme for the CM,
it differs from the latter in several important points.
First, for NCMs, the spin density response cannot be shown to be exactly decoupled from the charge response, i.e.,
there is no obvious transverse and longitudinal channel \cite{Buczek2020},
and, in Eq.~\eqref{eq:suscDysonEquation}, the Coulomb interaction must not be omitted.
In general, one expects the modes to be of mixed charge-spin character.
From the algorithmic perspective, one particular aspect of density response evaluation in NC systems turns out to become particularly challenging compared to the collinear case, namely the evaluation of the trace over spin indices $\alpha,\beta,\ldots$ in Eq.~\eqref{eq:chiKS}.
Up to 256 different matrix elements must be evaluated when the KKR form of the GF is deployed.
We manage this complexity by resorting to the automatic FORTRAN code generation based on a computational symbolic algebra.

Finally, there are several sum rules which the response function obeys and which are often not trivially reflected in the actual numerical calculations.
In the case of spin dynamics,
a fundamental one is the appearance of Nambu-Goldstone bosons (NGBs), i.e., 
the magnons of vanishing energy for $q = 0$.
NGBs emerge upon the spontaneous breaking of continuous symmetries, the magnetic ordering being an exemplary manifestation of the latter.
In the well known collinear case, the original symmetry of the Hamiltonian, SO(3), spontaneously breaks down to SO(2).
However, even in the familiar cases of ferro- and antiferromagnetic orderings yield two remarkably different pictures of this phenomenon.
As pointed out by Nambu \cite{Nambu2004}, in both cases there are two broken continuous symmetry generators but in ferromagnets they become canonically conjugated yielding only one NGB mode with quadratic dispersion relation.
On the contrary, in AFMs, the two corresponding NGBs remain independent and both feature linear dispersion.
The spontaneous symmetry breaking in the case of the frustrated AFMs is even more complex.
Here, the symmetry group of the magnetic system is closely related to the one of the rigid rotor. The ground state is characterized by three unconjugated generators of broken symmetries \cite{Pradenas2025} yielding three NGBs of linear dispersion \cite{Chernyshev2009}.
Through the deployment of a careful convergence control, our numerical scheme respects this symmetry and correctly yields three linear magnon branches in the $q \to 0$ limit.

\begin{figure*}
	\includegraphics[width=0.95\textwidth,trim={0 0 0 0},clip]{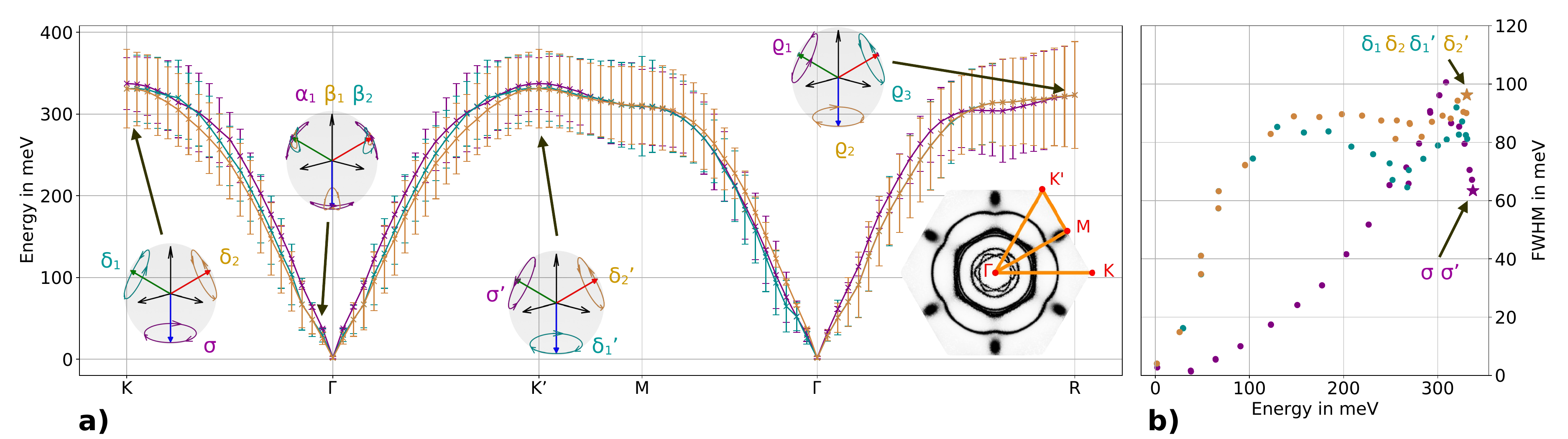}
	\caption{a) Magnon dispersion and inverse lifetimes (shown as bars designating the FWHM of the magnon peak) of Mn$_3$Rh obtained from our LRTDDFT calculations.
	The insets show the shapes of the magnon modes for selected momenta and a sketch of the (111)-cut of the Brillouin zone, combined with the corresponding Fermi surface.
	The precessing moments' trajectories are color-coded according to the dispersions of different modes.
	\textbf{b}) Magnons' FWHMs as a function of their energy, along the K-$\Gamma$-K' path.}
	\label{fig:magnonSpectrum}
\end{figure*}

\textit{Results -}
We proceed now to analysis of the spin dynamics of a prototypical KAFM, Mn$_3$Rh.
To start, we single out the features of the NC electronic band structure
essential for understanding the particularities of the Landau damping in this system.
Despite its vanishing net magnetization, the band structure of KAFMs is intrinsically spin-polarized, akin to the case of altermagnets, cf.\ Fig.~\ref{fig:bsf}.
(The directions within the Brillouin zone can be found as an inset in Fig.~\ref{fig:magnonSpectrum}.)
Additionally, Fig.~\ref{fig:bsf} depicts the Mn atoms' contributions to the total band structure.
Around $\Gamma$, one clearly distinguishes occupied electronic bands polarized along the respective moments' directions and responsible for their formation.
However, their location well below the Fermi level renders them inoperative in the spin-wave damping.
Concerning the formation of the non-collinear Stoner states responsible for the polarization-selective Landau attenuation, three other features are pivotal.
The strongly dispersing bands crossing the Fermi level (\enquote{O-states}) serve as a reservoir of the pairs' initial (occupied) electrons.
The intraband Stoner transitions within these bands are responsible for the damping of low-$q$ magnons.
Just above the Fermi energy, bands of $120$\textdegree{} rotated polarization appear close to the $K$ and $K'$ points
(we will refer to them as \enquote{T-states})
and turn out to be instrumental in the attenuation of the high momentum magnons.
Finally, there is an asymmetry of the spectral intensity of these states in the Brillouin zone, seen clearly, e.g., along the $K-\Gamma-K'$ directions for Mn2 and Mn3 atoms.
It is a consequence of the altermagnetic character of the band structure
as the amplitude of the Bloch states (electronic occupation and spin polarization) on a particular magnetic atom can differ even for wave vectors related by the symmetry of the underlying non-magnetic lattice.
In plain words, this altermagnetic behavior is caused by the fact that magnetic moments of different directions are embedded in different crystal environments imposed by the kagome pattern.

Now, we focus on the spin density fluctuations.
Contrary to the case of magnons in the transverse channel of the CM \cite{Buczek2020}, no general argument exists dictating that, in NCMs, there should be a class of density excitations of pure spin character for $q > 0$.
Nevertheless, we observe that in Mn$_3$Rh an \textit{effective transverse excitation subspace} forms (with its orientation varying necessarily in space).
It consists of modes involving practically only oscillating spin density oriented perpendicular to the local ground state magnetization direction
and we identify these modes as magnons.
We note that there emerge a further rich families of density excitations coupling charge and spin, including longitudinal spin fluctuations and plasmons,
however,
they appear above the magnon energy window and they will be studied elsewhere.
Fig.~\ref{fig:magnonSpectrum} presents the dispersion and damping of the magnons.
As expected from the preceding discussion of the spontaneous symmetry breaking, there are three distinct linearly dispersing magnon modes in the long wave-length limit. 
AT $K$ and $K'$, the modes reach their maximal energy of around 330 meV.
The dispersion agrees qualitatively with the adiabatic results \cite{leblancImpactFurtherrangeExchange2021} when the long-rage exchange interactions between moments are correctly accounted for 
which is automatically given in our \textit{ab initio} scheme.

The modes feature non-trivial spatial forms (also \textit{polarizations} or \textit{chiralities}).
Their variance between primitive cells is given given by the Bloch factor $e^{i\vec{q}\cdot\vec{R}}$ and within the cell by the eigenvectors of the loss matrix
of which examples are shown in Fig.~\ref{fig:magnonSpectrum}.
In general, in every magnon mode, all three NC moments undergo a complex coupled precession.
However, an interesting observation for the high symmetry points can be made.
At point $K$, for instance, two out of the three eigenmodes are energetically degenerate ($\delta$-modes and the separate $\sigma$-mode).
It turns out that the $\sigma$-mode is entirely localized at the Mn2 sublattice while the $\delta$-modes are supported on the Mn1 and Mn3 sublattices, respectively.
(For the symmetry related $M'$ point, the $\sigma$-mode is localized on the Mn2 sublattice, consistent with the crystal structure, cf. Fig.~\ref{fig:structure}b).)
This clear sublattice localization is additionally found for the triply degenerate out of 111-plane high symmetry point $R$ but it is not generally given for lower symmetry points in the Brillouin zone.


Let us now proceed to the discussion of the gripping picture of the system's Landau damping. 
The magnons in the entire BZ are well-defined
and representable as underdamped harmonic oscillators.
For small and intermediate momenta, the damping increases with the magnon energy,
as expected from the growing phase space available for Stoner excitations,
reaching the maximal value of 100meV given as full width at half-maximum (FWHM) of the magnon peak,
cf. Fig.~\ref{fig:magnonSpectrum}b).
An unexpected observation is that modes of comparable momenta and energies,
but different polarization feature strikingly different life-times.
For 100meV magnons, the difference exceed the factor of 4, as evident from Fig.~\ref{fig:magnonSpectrum}b. 
Similarly, the $\delta$-modes' damping at the high symmetry points $K$ and $K'$ is almost 50\% higher than that of the $\sigma$-mode.
Given the sublattice localization of these modes discussed just before, this paves the way to an effective spin-precession control on the atomic scale.
Finally, the high energy magnons show remarkable spintronic potential in Mn$_3$Rh due to the fact that their FHWM to frequency ratio is smaller than for their low energy counterparts.
Despite being more strongly damped, they undergo larger number of precessions before decaying.

\begin{figure}
	\includegraphics[width=\columnwidth]{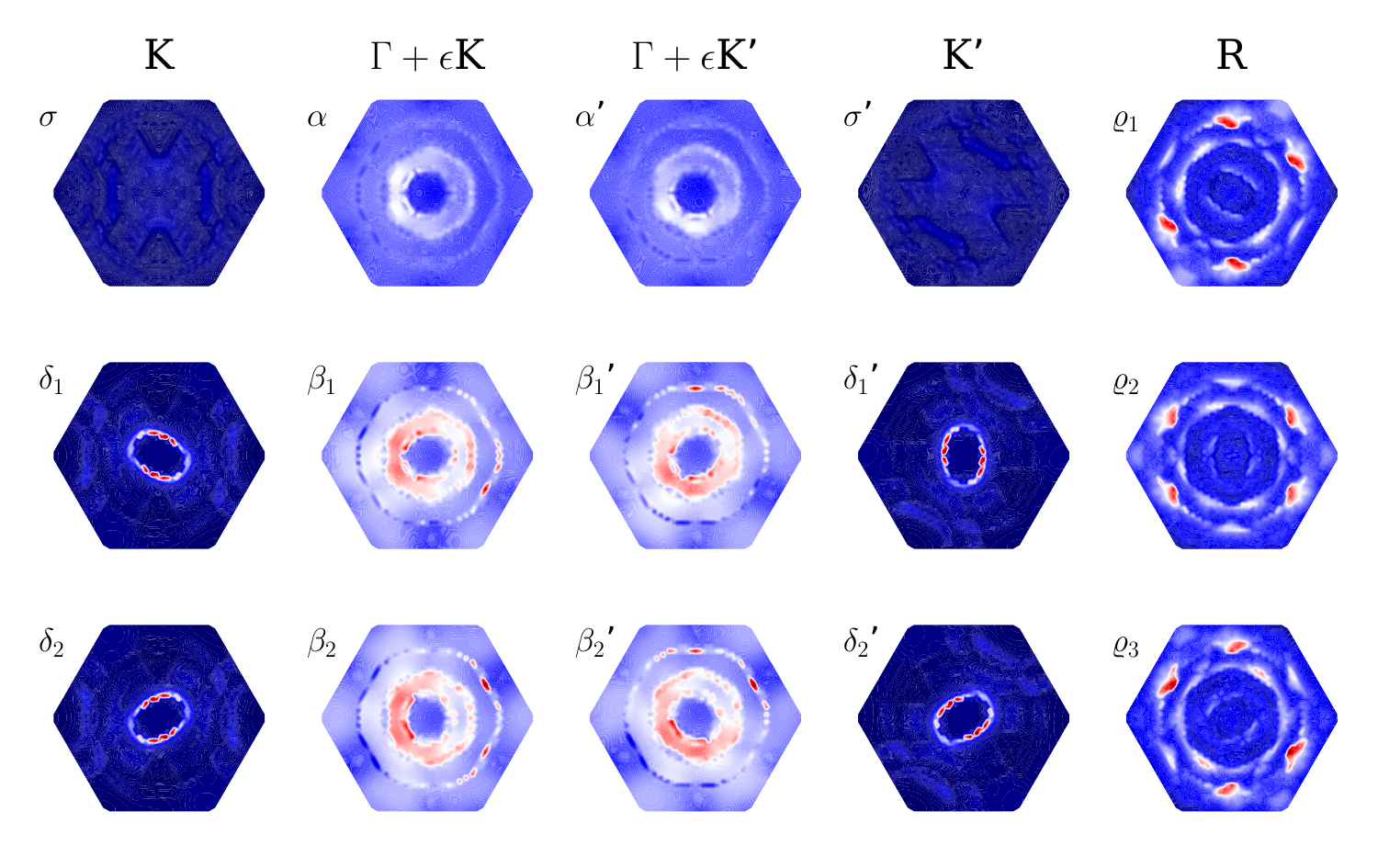}
	\caption{Landau maps for selected momenta (columns) and all three magnon polarization (rows). Discussion in the text.}
	\label{fig:LandauDamping}
\end{figure}

While analyzing this strong mode-selective life time difference, it is instructive to inquire which NC Stoner pairs are excited by the magnons of a given momentum $\vec{q}$ and cause the damping.
To this end, we use the concept of Landau maps, showing the density in momentum space of occupied electronic states with initial momentum $\vec{k}$ forming these Stoner pairs, cf. Fig.~\ref{fig:LandauDamping}.
The final states reside within bands above the Fermi level and feature momentum $\vec{k} + \vec{q}$.
As expected, for small momenta $q$ ($\Gamma + \epsilon K, \Gamma + \epsilon K'$), the maps resemble the form of the Fermi surface, cf. the inset of Fig.~\ref{fig:LandauDamping}.
The damping is of a clear intra-band character as the
small energy and momentum do not allow particle-hole pair formation across different bands crossing the Fermi level due to their significant Fermi velocity.
Nevertheless, it is evident that differently polarized magnons clearly favor transitions within specific electronic bands and momenta. 
This picture becomes even more intricate for increasing wave vectors
when the inter-band excitations start to play the decisive role.
In general, due to the large available phase space for the Stoner pair creation, one does not expect the maps to have hotspots, i.e., isolated momenta $\vec{k}$ in the BZ favored in the damping process.
Surprisingly, in the system under investigation, this depends strongly on the magnon polarization.
As shown in Fig.~\ref{fig:LandauDamping}, for the $K$ and $K'$ magnons, the two $\delta$-modes exhibit pronounced Landau hotspots and shortened life-time (FWHM of 97meV), whereas the $\sigma$-mode is damped by a more balanced distribution of Stoner excitations within the Brillouin zone and lives longer (FWHM of 62meV).
It appears that the hotspots and the resulting resonant Landau damping correspond to the Stoner pairs with an electron promoted from the O to the T states discussed above, cf.\ Fig.~\ref{fig:bsf}.
Because of the T-states depletion around $K$ on the Mn2 atom where the $\sigma$-mode is localized,
the hotspots do not form and the $\sigma$ magnons lives longer than its $\delta$ counterparts.
It is gratifying to observe the altermagnetic symmetry at work when focusing on the magnons at $K'$-point.
In this case, the longer living $\sigma$ magnon is localized on the Mn3 atom with reduced density of the T-states at $K'$ and the stronger damped $\delta$ magnons localized at atoms Mn1 and Mn2 where these states are available.


\textit{Summary -} Using our novel first principles computational scheme based on the \textit{ab initio} time-dependent density functional theory, we investigated the dispersion, spatial shapes, and the Landau damping magnons in non-collinear triangular kagome antiferromagnet Mn$_3$Rh.
We found the decay rates to differ strongly depending on the magnon polarization even for modes of comparable momentum and energy.
Magnons of certain polarization involve in the resonant Landau damping resulting in the hotspots in the Brillouin zone signifying that mostly electrons of specific momenta form the Stoner pairs the magnons decay into.
At the same time, there appear comparably long-living high energy modes of remarkable spintronic potential. 
We link this polarization-selective damping to the underlying non-collinear altermagnetic band structure.
We hope our study will stimulate further developments in the non-collinear magnonics.

\begin{acknowledgments}
We thank Leonid M. Sandratskii, L\'a{}szl\'o{} Szunyogh, Patrick Perndorfer, Sebastian Paischer, and Igor Maznichenko for fruitful discussions.
P.B. and A.E. acknowledge the funding, respectively, by \"Osterreichischer
Fonds zur F\"orderung der Wissenschaftlichen Forschung (FWF)
under grant I 5384 and by Deutsche Forschungsgemeinschaft (DFG) under grant BU 4062/1-1.
\end{acknowledgments}

\bibliography{magnetism.bib,paper_nc_tec.bib,altermagnets.bib}
\end{document}